\documentclass[prl,showpacs,superscriptaddress,preprint]{revtex4}%
\usepackage{amsfonts}
\usepackage{amsmath}
\usepackage{amssymb}
\usepackage{graphicx}%
\providecommand{\U}[1]{\protect\rule{.1in}{.1in}}

\begin{document}
\title{On the origin and universality of dislocation creation and void nucleation in FCC ductile metals}
\author{Wei-Wei Pang}
\affiliation{LCP, Institute of Applied Physics and Computational Mathematics, Beijing
100088, People's Republic of China}
\author{Ping Zhang}
\thanks{zhang\_ping@iapcm.ac.cn}
\affiliation{LCP, Institute of Applied Physics and Computational Mathematics, Beijing
100088, People's Republic of China}
\affiliation{Beijing Computational Science Research Center, Beijing 100089, People's
Republic of China}
\author{Guang-Cai Zhang}
\thanks{zhang\_guangcai@iapcm.ac.cn}
\affiliation{LCP, Institute of Applied Physics and Computational Mathematics, Beijing
100088, People's Republic of China}
\author{Ai-Guo Xu}
\affiliation{LCP, Institute of Applied Physics and Computational Mathematics, Beijing
100088, People's Republic of China}
\author{Xian-Geng Zhao}
\affiliation{LCP, Institute of Applied Physics and Computational Mathematics, Beijing
100088, People's Republic of China}


\begin{abstract}
We clarify via molecular dynamic simulations and theoretical analysis the origin of dislocation creation and void nucleation during uniaxial tensile process in face-centered-cubic (FCC) ductile metals. We show that the dislocations are created through three distinguished stages: (i) Flattened octahedral structures (FOSs) are randomly activated by thermal fluctuations; (ii) The double-layer defect clusters are formed by self-organized stacking of FOSs on the close-packed plane; (iii) The stacking faults surrounded by the Shockley partial dislocations are created from the double-layer defect cluster due to the relative slip of internal atoms. Whereas, the void nucleation is shown to follow a two-stages description: (i) The vacancy strings are first formed by intersection of different stacking faults; (ii) Then the vacancy strings transform into the voids by emitting dislocations. We demonstrate that our findings on the origin of dislocation creation and void nucleation is universal for a variety of FCC ductile metals with low stacking fault energy.
\end{abstract}
\maketitle

The stretch loading and shock unloading damage processes of ductile metals involve complicated generation and evolution of a series of microscopic structures such as dislocations and voids \cite{Seaman, Lubarda, Seppala}, a deep knowledge of which compose the most fundamental basis for a predictive dynamic fracture modeling. Under low and mediate strain rates, it is generally believed that due to the fact that the material is mechanically near equilibrium and experiences a process of low energy state, these microscopic structures are created in the energetically activated regions (grain boundary, for instance) manifested by atomic deviations from ideal crystal lattice sites. Under high strain rate, however, because the inertia plays a dominant role, there is no time enough to release the local stresses. In this case, striking mechanical non-equilibrium drives the material to experience various excited high energy states. As a response, the dislocations and voids can be generated in the interior of bulk crystal with the help of thermal fluctuations.

To describe and understand the dislocation and void nucleation processes that the present experimental advance are extremely difficult to catch the microscopic mechanism, numerous theoretical efforts, mainly through molecular dynamics simulations, have been paid in the last decade. Void surface \cite{Takahiro}, crack tip \cite{Berk}, energetic atomic clusters with larger relative displacements \cite{Zuo}, bicrystal interfaces \cite{Douglas}, grain boundaries \cite{Swygenhoven}, as well as free surfaces \cite{Liu,Zhu}, have been shown to provide reliable activated volumes for dislocation nucleation driven by critical local shear stress. Most of these studies were mainly focused on the conditions and influential factors responsible for dislocation nucleation with the aim to provide explanative clues to understand critical yield phenomena and the corresponding microscopic plasticity. Whereas, to date an essential physical picture on the self-organized atomic collective motions during dislocation nucleation keeps unclear. The reason is that no effective description of universal sense has been given to classify the complicated atomic configurations of dislocations within the activated volume. For void nucleation, it has been acknowledged that under low strain rate or quasi static stretching, voids occur predominantly via the second-phase particle cracking or second-phase debonding from the matrix material \cite{Srinirasan}. Under the moderate stretching, voids nucleate preferentially at the grain boundary junctions and grow along the grain boundary \cite{Preston,Dongare}. Under high strain rate, various microscopic or mesoscopic voids may evolve both inside the bulk crystal \cite{Belak} and on the grain boundaries \cite{Yuan}. These results prove to be indispensable for understanding the onset and development of dynamic damage. Again, however, due to the very scarce knowledge on the extreme diversity in structural configurations around the void nucleation core, a general description on the shapes and distribution of nucleated voids, thus the reliable mechanisms for void nucleation, are still totally lacking.

Inspired by the above-mentioned observation, in the present letter, as a first step we try to understand the atomic self-organized collective motion law for the dislocation creation and clarify the mechanisms for the void nucleation at high strain rate. For this aim, in our molecular dynamics simulations we choose to select single-crystal FCC metals, in which because the atoms are in order, the newly created microscopic structures are possible to be identified and tracked, and because the stress is uniform, the creation mechanism for the microscopic structures are feasible to be analyzed. We give a three-stage picture on the dislocation nucleation: (i) The FOSs (see below for definition) are firstly activated by thermal fluctuations; (ii) The defect clusters are formed by self-organized stacking of FOSs on the close-packed plane; (iii) The stacking faults are formed by the slip of atoms within the defect clusters. We also propose a two-stage mechanism for the void nucleation: (i) The vacancy strings are firstly formed by intersection of two stacking faults; (ii) Then the vacancy strings transform into the voids by emitting dislocations. We show that these general findings can be applied to a variety of FCC ductile metals with low stacking fault energy.

The materials we use for simulations include Ag, Au, Cu, Ni, Pt, and Pd. These metal have low stacking fault energy. The simulation tool is the well-known LAMMPS software package \cite{LAMMPS}. The interatomic interaction is described by an embedded atom method (EAM) potential \cite{Daw,Foiles}. The simulation box consists of $80\times80\times80$ unit cells and contains approximately $2\times10^{6}$ atoms. Periodic boundary conditions are used to minimize surface and edge effects. The system is initially equilibrated at temperature $T=3$ K and ambient pressure $P=0$ GPa. Once the equilibrium is established, the thermostat is turned off and uniaxial tensile strain is applied along $[100]$ direction with two constant strain rates $\dot{\varepsilon}=10^{9}/s$ and $10^{8}/s$ for comparison. The atoms are distinguished by calculating their coordination numbers and common neighbor analysis (CNA) values.

\begin{figure}[ptb]
\begin{center}
\includegraphics[width=0.7\linewidth]{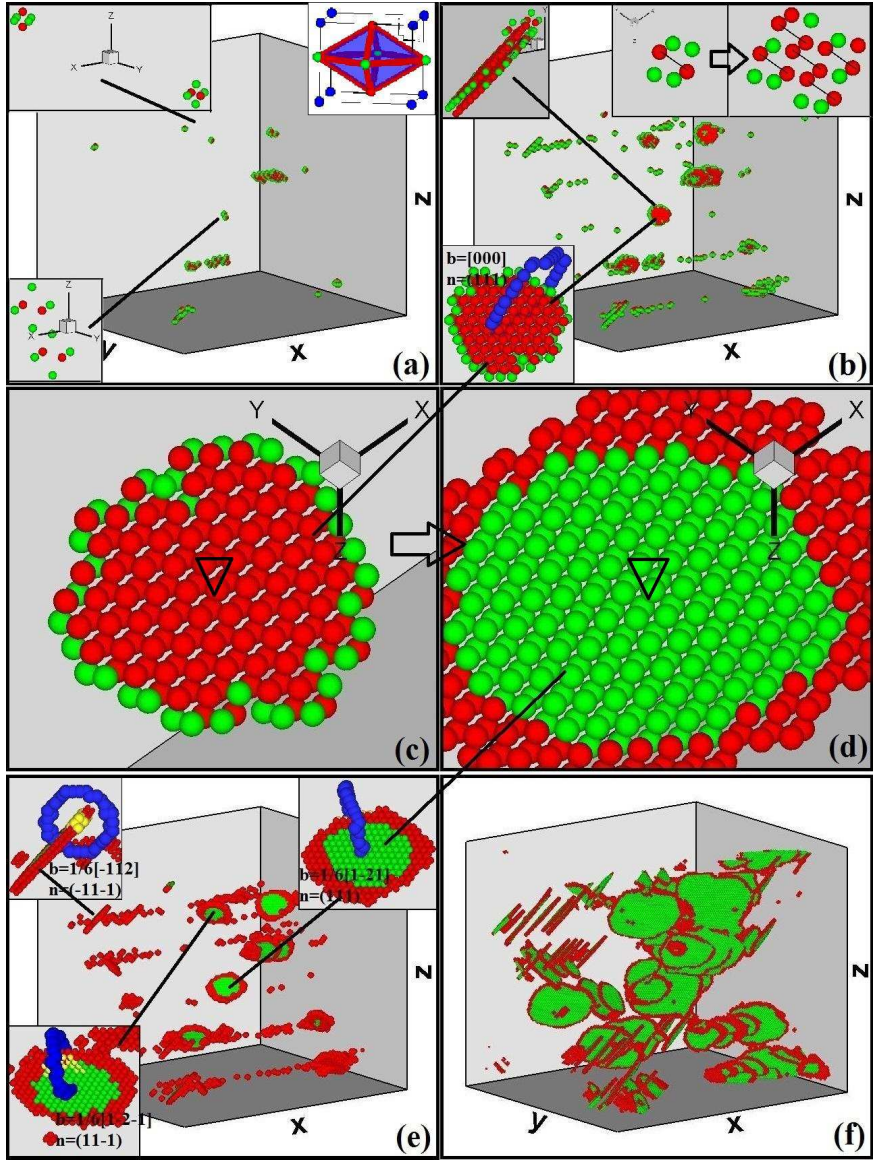}
\end{center}
\caption{Molecular-dynamics simulation snapshots that provide a general three-stage physical picture for the generation of dislocations and the corresponding non-zero Burgers vectors in FCC ductile metals under high-strain-rate uniaxial strech. Panel (a) shows that FOSs (for a detailed view, see the top right inset) are firstly activated in the metals by thermal fluctuations. Panel (b) shows that FOSs begin to stack on the close-packed plane to form double-layer defect clusters (see the top left inset for closer view). This stacking process is shown in the top right inset. The Burgers vector for the double-layer defect cluster structure is calculated to be zero, as shown in the bottom left inset. Panel (c) and panel (d) shows the transformation of the double-layer defect clusters into stacking faults. Panel (e) gives a few non-zero Burgers vectors of the nucleated dislocations that surround the stacking faults. Panel (f) shows the growth of stacking faults and dislocations. In panels (a)-(c)the coordination numbers of red and green atoms are 13 and 12, respectively, while in panels (d)-(f) the CNA values of red and green atoms are 5 and 2, respectively}%
\label{defect}%
\end{figure}

During the early stage of loading, the system responds elastically and the lattices are stretched without dislocations formed. As the strain increases, some clusters of atoms deviate from their equilibrium lattice positions, as shown in Fig. \ref{defect}(a). Strikingly, most of these atomic clusters form what we henceforth call flattened octahedral structures (FOSs). The six vertex atoms of FOS are the face-centered atoms of the FCC crystal, see the top right inset in Fig. \ref{defect}(a). The formation of FOS can be understood as follows: Under the tensile loading along the $x$ direction, to remain the minimum energy state, two face-centered atoms along the $z$ or the $y$ axis become closer. The shifts of these two atoms lead to that they become neighboring atoms and simultaneously change the CNA values of other four atoms. Therefore, the FOS can be identified by the CNA value. When the strain is small, the FOSs are sparse and randomly distributed. The FOSs are dynamical in that they may stochastically occur, annihilate, and reoccur in other places.

With increasing the stress, in the zones with cumulative FOSs, some FOSs tend to stack on the close-packed plane to form double-layer defect clusters, see Fig. \ref{defect}(b) and the top left inset inside. This stacking process can be understood as follows: The FOSs attract with each other in tensile FCC metal, leading to the atoms other than the collapsed atoms of the FOSs to also collapse to form more FOSs that stack on the same close-packed plane, see the top right inset in Fig. \ref{defect}(b). We calculate the Burgers vector of the double-layer defect clusters according to the Frank scheme \cite{Wang}, see the bottom left inset in Fig. \ref{defect}(b), in which the blue loop is used to calculate the Burgers vector. The Burgers vector is determined to be $\bold{b}=[000]$, which demonstrates that the stacked FOSs initially form defect clusters other than stacking faults, and therefore no dislocations form at this FOS stacking stage.

When the size of the double-layer defect cluster exceeds a certain value, the central-region atoms of the cluster undergo a relative inter-layer slip. Such a process can be observed from two successive snapshots, Fig. \ref{defect}(c) and Fig. \ref{defect}(d), which are plotted in the way that one sees from the norm direction of close-packed plane. Before inter-layer slip [Fig. \ref{defect}(c)], one can see that the two layers are mostly overlapped, leaving large slits (see the white region surround by black triangle for one slit). In Fig. \ref{defect}(d), whereas, in the central region each slit is split into two little holes, while in the surrounding region, the slits remain the shape. This fact indicates that the central atoms in the  double-layer defect cluster are more active than the atoms in the surroundings. We calculate the Burgers vector of the slipped  double-layer defect cluster shown in Fig. \ref{defect}(d), and find  $\bold{b}=[1\bar{2}1]$ [see the top right inset in Fig. \ref{defect}(e)]. The non-zero Burgers vector demonstrates that the dislocations are generated from the double-layer defect cluster due to the relative slip of the internal atoms.

After systematic calculations on the Burgers vectors of  the nucleated dislocations in the simulation block, we find that these dislocations are all Shockley partial dislocations, such as $(111)-\frac{1}{6}[1\bar{2}1]$ and $(\bar{1}1\bar{1})-\frac{1}{6}[\bar{1}12]$ shown in the inset in Fig. \ref{defect}(e). It is well known that generally there are at most twelve types of Shockley partial dislocations to possibly appear in tensile FCC metals. Strikingly, in our simulations we find that four types of partial dislocations, namely, $(111)-\frac{1}{6}[211]$, $(1\bar{1}1)-\frac{1}{6}[2\bar{1}1]$, $(11\bar{1})-\frac{1}{6}[21\bar{1}]$, and $(\bar{1}11)-\frac{1}{6}[\bar{2}11]$, fail to occur. The reason is ultimately due to that under the present tensile loading along the [100] direction, no atoms along the [100] direction are collapsed to form the FOSs. Once small dislocations occur, they would grow up quickly and multiply to cover the whole simulation block, as shown in Fig. \ref{defect}(f).

\begin{figure}[ptb]
\begin{center}
\includegraphics[width=0.8\linewidth]{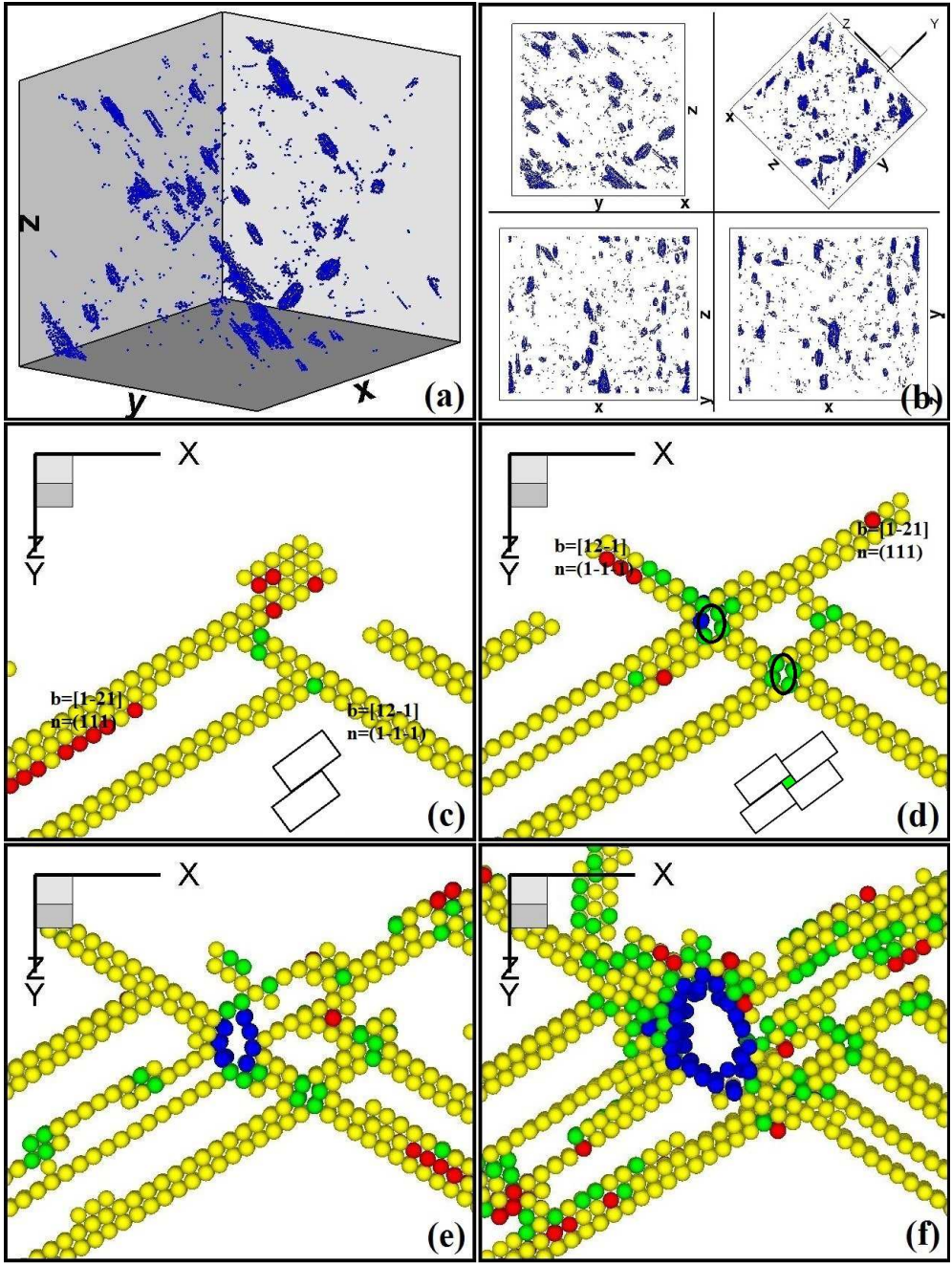}
\end{center}
\caption{Incipient void nucleation phenomenon and its two-stage mechanism. Panel (a) and (b) shows the void nucleation phenomenon. Panel (c) shows that four stacking faults appearing as lines nucleate from double-layer defect clusters. Panel (d) shows that two pillar-like vacancy strings are generated from the the intersections of stacking faults, see the upper and lower black circles. Panel (e) shows that the upper vacancy string transforms into a void via emitting dislocations, while the lower one retains its size. Panel (f) shows that nucleated voids grow gradually and neighboring vacancy strings disappear.}%
\label{void nucleation}%
\end{figure}

The dislocation nucleation and slip release part of the shear stress, but dot not release bulk stress (negative pressure). As a result, with increasing the tensile strain, plenty of energies accumulate in the system. To release energy, some voids or cracks may be generated at the weak points in material. This is what we observe that the strain ultimately drives some voids to nucleate in dislocation aggregation regions, which are generally considered as weak points.
Figure \ref{void nucleation}(a) shows the nucleated voids inside the simulation box. When loading along the $[100]$ direction, the nucleation of voids is random inside the simulation box and most incipient void shapes are found to be pillar-like. Figure \ref{void nucleation}(b) shows the nucleated voids from different views. Peculiarly, we find that the elongations of voids are predominantly along the $[011]$ and $[01\bar{1}]$ directions that are vertical to the loading direction, while the voids of other directions do not grow up.

To further reveal the incipient nucleation mechanism of voids, we picked out a layer, with a thickness of $2.3$ nm, perpendicular to the elongation direction
of void. Figures \ref{void nucleation}(c)-(f) show the evolutionary process of these atoms in the layer, where only the defect atoms are shown and the direction
of principal plane is $[0\bar{1}1]$. Specially, Fig. \ref{void nucleation}(c) shows four stacking faults which appear as lines in the present view. The left two stacking faults have the normal direction $[111]$ and the Burgers vector $\frac{1}{6}[1\bar{2}1]$, while the right two ones have the normal direction $[1\bar{1}\bar{1}]$ and the Burgers vector $\frac{1}{6}[12\bar{1}]$. These stacking faults grow to be larger with time under the tensile loading. Meanwhile, stacking faults with different normal directions evolve to intersect with each other and generate pillar-like vacancy strings located at the intercrossing lines. Figure \ref{void nucleation}(d) shows two vacancy strings (indicated by two black circles) resulting from the intersections of the left two and the right one stacking faults.
These vacancy strings can grow into voids, provided dislocations are emitted from them. However, to emit dislocations, the size of vacancy string and the stress around the vacancy string must exceed some critical values. In Fig. \ref{void nucleation}(e) we could observe that the upper vacancy string grows up into a void via emitting dislocations, while the lower one retains its size. This is because there are more activated atoms around the upper vacancy string. The release of stress resulting from the growth of the nucleated voids will suppress the growth of neighboring vacancy strings. As a result, one can see from Fig. \ref{void nucleation}(f) that the nucleated void further grow and the other vacancy strings tend to disappear. In addition, We can also observe that the void shape gradually evolves from pillar-like into ellipsoidal.

The above process of vacancy string creation via the intersection of two stacking faults could be regarded as two successive plastic deformations. The
first deformation brings a stacking fault into the system. The atoms have a displacement of the corresponding Burgers vector along the plane. During the second deformation process, the atoms further have a corresponding displacement along the other plane. The plastic deformation resulting from stacking faults can be
described by a distortion tensor
$\mbox{\boldmath$\beta$} =\mbox{\boldmath$\delta$}(\Sigma)\mathbf{b} = \int\int_{\sum}d\mathbf{s^{\prime}\delta(r^{\prime}-r)b}$,
where $\mbox{\boldmath$\delta$}(\Sigma)$ is the surface Dirac function, $\Sigma$ is stacking fault plane, and $\mathbf{b}$ is the Burgers vector.
The relative volume variation is $\delta V/V = \mbox{Tr (\boldmath $\beta$)}$.
For the case of single stacking fault,
$\mbox{Tr ({\boldmath$\beta$})} = \mbox{\boldmath$\delta$}(\Sigma)\mathbf{\cdot b} = 0$, therefore, there is no density variation in the system.
For the case of two stacking faults intersecting with each other,
the distortion tensor is
$\mbox{\boldmath$\beta$} = \mbox{{\boldmath$\beta$}$_1$+{\boldmath$\beta$}$_2$}\cdot(\mathbf{I}+\mbox{{\boldmath$\beta$}$_1$})$,
where \mbox{{\boldmath$\beta$}$_{1}$} and \mbox{{\boldmath$\beta$}$_{2}$} are the distortion tensors of the two stacking faults, respectively.
The volume variation is
$\delta V = \mathrm{\int} dV \mbox{Tr ({\boldmath$\beta$}$_1 \cdot$ {\boldmath$\beta$}$_2$)} =(\mathbf{b}_{2}\cdot\mathbf{n}_{1})(\mathbf{b}_{1}\cdot\mathbf{n}_{2})L/|\mathbf{n}_{1}\times\mathbf{n}_{2}|$, where $\mathbf{n}_{1}$ and $\mathbf{n}_{2}$ are norm directions of the stacking faults and $L$ is the length of the vacancy string. Therefore, we arrive at that the cross-section area of vacancy string resulting from the intersection of two different stacking faults is $(\mathbf{b}_{2}\cdot\mathbf{n}_{1})(\mathbf{b}_{1}\cdot\mathbf{n}_{2})/|\mathbf{n}_{1}\times\mathbf{n}_{2}|$, and the direction of vacancy string is $\mathbf{n}_{1}\times\mathbf{n}_{2}$.

Since we have made it clear that the dislocations generated from the double-layer defect clusters are all Shockley partial dislocations, and the corresponding Burgers vectors have been obtained, then according to the above expression, we can determine that the initial vacancy strings have a typical cross-section area of $\sqrt{2}a^2/36$ ($a$ is the lattice constant), and particularly, their distribution directions have six possible types. However, in Fig. \ref{void nucleation}(a) for voids evolved from vacancy strings, we only observe two types of voids with respective directions of $[011]$ and $[0\bar{1}1]$, which are vertical to the loading direction. This phenomenon can be qualitatively explained as follows: The energy released via growth of a vacancy string can be expressed as $\int\int d \mathbf{s} \cdot \mbox{\boldmath$\sigma$}  \cdot \delta \mathbf{r}$,
where $\mbox{\boldmath$\sigma$}$ is the applied stress, $\mathbf{s}$ is the surface area of the vacancy string, and $\delta\mathbf{r}$ is the growth displacement. Since in our simulation setup the applied stress is $\sigma_{xx}$. Thus, only when the vacancy string is vertical to $[100]$, which enables it to have larger projective area along $[100]$, does its growth release more energy and evolve into voids. This analysis well explains the numerical results shown in Fig. \ref{void nucleation}(a).

\begin{figure}[ptb]
\begin{center}
\includegraphics[width=0.8\linewidth]{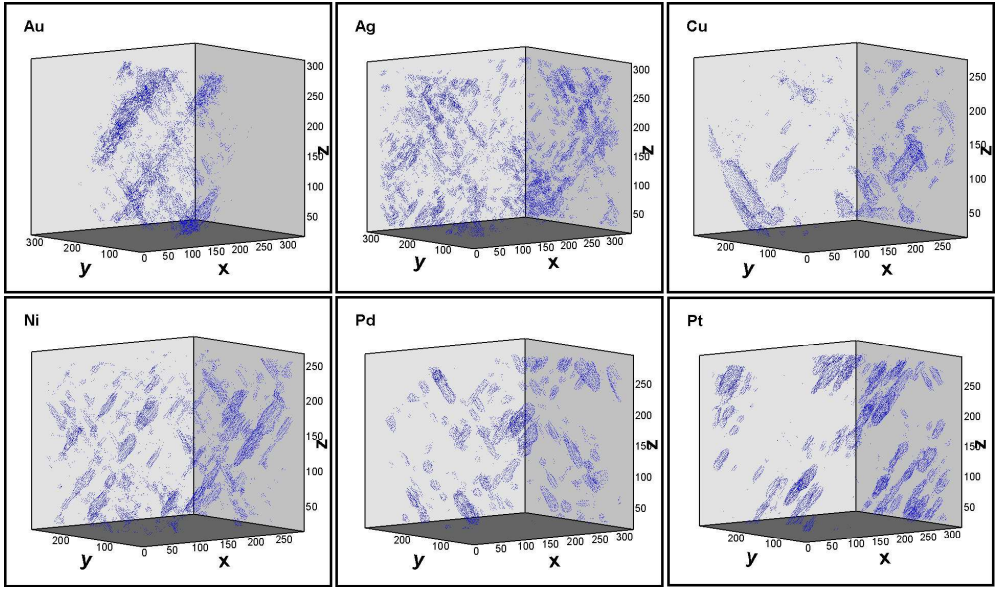}
\end{center}
\caption{Molecular-dynamics simulation results for shapes and distributions of nucleated voids in six different ductile metals. }%
\label{nucleation process}%
\end{figure}

We turn now to demonstrate that our present findings on the dislocation creation and void nucleation at high strain rate are universal for a variety of tensile FCC metals. For this purpose, we have simulated six types of FCC metals as mentioned above. The typical results are shown in Fig. \ref{nucleation process}, from which one can clearly see that although the number densities of nucleated voids are different in different metals, the nucleation processes show similar behaviors. Specially, we have verified that the initial FOS generation and stacking, dislocation creation and the corresponding Burgers vectors, vacancy string formation, as well as void shape and direction, are all analogous for the six metals. For example, as revealed in Fig. \ref{nucleation process}, under loading along $[100]$ direction most voids in all the six metals show pillar-like shapes, and their elongations are predominantly along the directions $[011]$ and $[01\bar{1}]$.

As a final remark, it should be noticed that all the materials under investigation in this work have low stacking fault energy. In these materials, the partial dislocations and stacking faults are initiated during the plastic procedure. Whereas, for materials with high stacking fault energy, for example, the aluminum,
initially created dislocations are almost perfect dislocations and nearly no stacking faults can be formed during the plastic procedure. In such materials, tensional loading with high strain rate generally leads to point defects or microcracks, instead of the voids. In a word, the newly found mechanisms for dislocation creation and void nucleation in the present paper are valid for ductile metals with low stacking fault energy.

In conclusion, through systematic molecular dynamics simulations and a rationalized analysis on the evolution behavior of several typical FCC ductile metals under high-strain-rate uniaxial tension, we have provided a general physical picture for the dislocation creation and void nucleation. We have shown that the dislocation creation follows a three-stage procedure, in which random FOSs are at first activated by thermal fluctuations, then the FOSs form double-layer defect clusters via stacking on the close-packed planes, and finally these double-layer defect clusters evolve into Shockley  partial dislocations due to relative slip of internal atoms. Whereas, the void nucleation follows a two-stage procedure, in which the first stage is characterized by the generation of pillar-like vacancy strings through intersections of different stacking faults, while the second stage is represented by transformation of vacancy strings vertical to the loading direction into voids via emitting dislocations. Our findings are expected to pave a way to build up a universal understanding on the origin of dislocation creation and void nucleation in a variety of ductile metals, which we believe is fundamental for accurate dynamic damage fracture modeling.

We acknowledge support of the National Natural Science Foundation of China under Grants No. 51071032, No. 11075021, and No. 11102026, and Science Foundation of CAEP under
Grants No. 2011A0301016 and No. 2012B0101014.

\end{document}